\documentclass[reprint,onecolumn,showpacs,preprintnumbers,superscriptaddress]{revtex4-1}
\usepackage[utf8]{inputenc}
\usepackage{epsfig}
\usepackage{epstopdf}
\usepackage{amsmath,amssymb,amsthm}
\usepackage{dcolumn}
\usepackage{bm}
\usepackage{MnSymbol}
\usepackage{graphicx}
\usepackage{subfig}
\usepackage[justification=RaggedRight]{caption}

\usepackage{hyperref}
\usepackage{stmaryrd}

\usepackage{natbib}

\SetSymbolFont{stmry}{bold}{U}{stmry}{m}{n}
\begin{document}

\title{Turbulent transport of alpha particles in tokamak plasmas}
\author{A. Croitoru}
\email{c.m.andreea@gmail.com}
\affiliation{National Institute of Laser, Plasma and Radiation Physics, PO Box MG 36,
RO-077125 M\u{a}gurele, Bucharest, Romania}
\affiliation{Faculty of Physics, University of Bucharest, Romania}
\author{D. I. Palade}
\affiliation{National Institute of Laser, Plasma and Radiation Physics, PO Box MG 36,
RO-077125 M\u{a}gurele, Bucharest, Romania}
\affiliation{Faculty of Physics, University of Bucharest, Romania}
\author{M Vlad}
\affiliation{National Institute of Laser, Plasma and Radiation Physics, PO Box MG 36,
RO-077125 M\u{a}gurele, Bucharest, Romania}
\author{F Spineanu}
\affiliation{National Institute of Laser, Plasma and Radiation Physics, PO Box MG 36,
RO-077125 M\u{a}gurele, Bucharest, Romania}
\keywords{Turbulence, transport, alpha particle loss}

\begin{abstract}
We investigate the $\boldsymbol{E}\times \boldsymbol{B}$ diffusion of fusion born $\alpha $ particles
in tokamak plasmas. We determine the transport regimes for a realistic model
that has the characteristics of the ion temperature gradient (ITG) or of the trapped electron modes (TEM) driven
turbulence. It includes a spectrum of potential fluctuations that is modeled
using the results of the numerical simulations, the drift of the potential
with the effective diamagnetic velocity and the parallel motion. Our
semi-analytical statistical approach is based on the decorrelation
trajectory method (DTM), which is adapted to the gyrokinetic approximation.
We obtain the transport coefficients as a function of the parameters of the
turbulence and of the energy of the $\alpha $ particle. According to our results,
significant turbulent transport of the $\alpha $ particles can appear only
at energies of the order of 100KeV. We determine the corresponding
conditions.
\end{abstract}

\maketitle

\section{Introduction}

The turbulent transport of the fast $\alpha $ particles was considered
negligible in tokamak plasmas \cite{zweben2000alpha} due to the fast gyration motion
with a Larmor radius much larger than the correlation length, which leads to a
very small amplitude of the gyro-average potential. However, this problem
was reconsidered in the last decade \cite{zhou2011dependence,zhou2010turbulent,
dewhurst2010finite,heidbrink2009evidence,angioni2009gyrokinetic,hauff2006turbulent,hauff2008mechanisms,angioni2008gyrokinetic,
chowdhury2012nature,zhang2010scalings, zhang2008transport,gunter2007interaction,
gingell2014plasma}, in preparation of the Tritium experiments in JET and ITER. The conclusions are rather
dispersed, from completely negligible turbulent transport \cite{pace2013energetic} to
diffusion coefficients that can be larger than those of plasma ions \cite{estrada2006turbulent,vlad2005turbulent}.

Most of the theoretical studies of $\alpha$ particle turbulent transport are
self-consistent numerical simulations of turbulence and $\alpha $ particles
turbulent fluxes. Other studies use the test particle approach in numerical
simulations based on constructed potentials or on the results of turbulence
simulations. The characteristics of turbulence and the $\alpha $ particle
fluxes are determined in the first case as functions of the macroscopic
conditions (gradients, heating power). The second approach obtains the
diffusion coefficient of the $\alpha $\ particles as a function of their
energy and of the characteristics of the turbulence. It allows to find if
the turbulent loss of $\alpha $ particles can be significant and to identify
the corresponding conditions. 

This paper is included in the last category. We determine the transport
regimes of the $\alpha $ particles as a function of their energy for a
realistic model of turbulence. The spectrum of the stochastic potential has
the shape of the ion temperature gradient (ITG) or of the trapped electron modes (TEM) driven turbulence. The drift of the potential with the effective diamagnetic velocity and the parallel
motion of the fast particles are included in the model. The main result
consists in the evaluation of the energy corresponding to the maximum
turbulent transport $D_{\max }.$ We show that $D_{\max }$\ does not always
appear when the $\alpha $ particles reach the energy of the plasma ions (for the
ashes), but it can correspond to larger energies. 

We use a semi-analytical approach based on the decorrelation trajectory
method (DTM) \cite{vlad1998diffusion,vlad2004trajectory} in the gyrokinetic approximation
developed in \cite{hauff2006turbulent}.

The article is organized as follows. Section \ref{model} contains the basic
equations and the statistical approach based on the DTM for the Lorentz
transport in the gyro-kinetic approximation for a Maxwellian distribution of
Larmor radii. In section \ref{results} we present the transport regimes for
different ranges of the parameters and obtain the energy dependence of the $\alpha$ particles diffusion coefficient. The conclusions are
summarized in section \ref{conclusions}.

%ce aducem in plus: -"Introducerea miscarii paralele $dz/dz=v_{th}$ (si a unei valori finite a lungimii de corelatie a turbulentei) pentru a se studia dependent de energie a difuziei particulelor alpha din reactia de fuziune. Adica vrem sa vedem pe la ce energie sunt pierdute aceste particule (la ce energie este maxima difuzia in plasma turbulenta). Miscarea paralela a fost luata in considerare in [2], dar cu alt scop (dependent de raportul $v_\perp/v_\parallel$). Nu am facut atunci variatia energiei".

\section{Model and statistical method}

\label{model}

\subsection{Equations of motion for fast ions}

The turbulent transport of the fast ions is studied in the test particle
approach starting from the Newton-Lorentz equations of motion in a
stochastic potential $\phi (x_1,x_2,z,t)$ and a constant magnetic field $\mathbf{%
B}=B\mathbf{e}_{z}$ oriented along the $z$ axis

\begin{equation}
\begin{split}
\frac{du_{i}(t)}{dt}& =-\frac{q}{m_{\alpha }}\frac{\partial \phi (x_1,x_2,z,t)}{%
\partial x_{i}}+\Omega _{\alpha }\varepsilon _{ij}u_{j}, \\
\frac{dx_{i}(t)}{dt}& =u_{i}, \\
\frac{dz(t)}{dt}& =u_{z}.
\end{split}
\label{Lorentz2}
\end{equation}%
Where $\mathbf{x}(t)=(x_{1}(t),~x_{2}(t))$ is the ion
displacement in the plane perpendicular to $\mathbf{B}$, $u_{i},i=1,2$ are
the components of the velocity in this perpendicular plane, $u_{z}$ is the
velocity along the magnetic field, $m_{\alpha }$ is the ion mass, $q$ is its
charge, $\Omega _{\alpha }=qB/m_{\alpha }$ is the cyclotron frequency and $%
\varepsilon _{ij}$\ is the antisymmetric tensor\ ($\varepsilon _{12}=1,$ $%
\varepsilon _{21}=-1,$ $\varepsilon _{ii}=0)$.

The potential $\phi (x_{1},x_{2},z,t)$ is modeled as a stationary and homogeneous Gaussian stochastic
function. Its spectrum $%
S(k_{1},k_{2},z,t)$ corresponds to the general characteristics of the
ITG or TEM turbulence. Since the spectrum at saturation is mainly determined
by the ion dynamics, its shape is similar for both types of turbulence. The only
difference is given by the value of the typical wave numbers ($k_{2}\rho _{ion}\lesssim
1$ for ITG, $k_{2}\rho _{ion}\approx 1$ for TEM). The spectrum has two symmetrical
maxima for $k_{2}=\pm k_{0},$ $k_{1}=0,$ and zero amplitude for $k_{2}=0.$
We use the simple analytical expression of $S(\mathbf{k})$\ that was found
in \cite{vlad2015electron} to be in agreement with numerical and experimental
results of \cite{hauff2007b,
shafer20122d}

\begin{eqnarray}
S(k_{1},k_{2},z,t) &\propto &\Phi ^{2}\exp \left( -\frac{|z|}{
\lambda _{z}}-\frac{|t|}{\tau _{c}}\right)  \label{spectr} \\
&&\frac{k_{2}}{k_{0}}\exp \left( -\frac{k_{1}^{2}}{2}\lambda _{1}^{2}\right) %
\left[ \exp \left( -\frac{(k_{2}-k_{0})^{2}}{2}\lambda _{2}^{2}\right) -\exp
\left( -\frac{(k_{2}+k_{0})^{2}}{2}\lambda _{2}^{2}\right) \right] .  \notag
\end{eqnarray}%
The parameters of this function are the amplitude of the potential
fluctuations $\Phi ,$ the correlation lengths along each direction $\lambda
_{i},$ $i=1$ (radial), $i=2$ (poloidal), $i=z$ (parallel), and the
correlation time $\tau _{c}.$ The Fourier transform of $%
S(k_{1},k_{2},z,t)$\ is the Eulerian correlation (EC) of the potential.\ 

\bigskip The change of coordinates $(\mathbf{x},\mathbf{u})\longrightarrow ({%
\ \boldsymbol{\xi }},E,\mu ,\zeta )$ leads to

\begin{align}
\frac{d\xi _{i}}{dt}& =-\varepsilon _{ij}\frac{1}{B}\frac{\partial \phi (%
\boldsymbol{\xi }+\boldsymbol{\rho },z,t)}{\partial \xi _{j}},  \label{csi}
\\
\frac{d\rho _{i}}{dt}& =\varepsilon _{ij}\left[ \frac{1}{B}\frac{\partial
\phi (\boldsymbol{\xi }+\boldsymbol{\rho },z,t)}{\partial \xi _{j}}+\Omega
_{\alpha }\rho _{j}\right] ,  \label{rhovar} \\
\frac{dz}{dt}& =u_{z},  \label{parvar}
\end{align}%
where the time dependent Larmor radius $\boldsymbol{\rho }(t)$ is defined by $%
\rho _{i}=-\varepsilon _{ij}u_{j}/\Omega _{\alpha },$ $\boldsymbol{\xi }(t)$ is
the guiding center position $\boldsymbol{\xi }=\mathbf{x}-\boldsymbol{%
\rho }$, $E$ is the particle energy, $\mu =u_{\perp }^{2}/2B$ is the
magnetic moment, and $\zeta $ is the gyrophase angle.\ 

The very large value of the cyclotron frequency $\Omega_\alpha \gg 1$ enables a strong
simplification of the equation of motion by using the gyrokinetic
approximation \cite{gyrokin2,gyrokin1}. The first term in the right hand side term of Eq. (\ref%
{rhovar}) is negligible compared to the second one, and the solution of this
equation is $\boldsymbol{\rho }(t)=\rho_0\left(\sin (\zeta _{0}+\Omega _{\alpha
}t),\cos (\zeta _{0}+\Omega _{\alpha }t)\right)$, where $\rho_0=u/\Omega
_{\alpha }$\ and $u=\sqrt{u_{1}^{2}+u_{2}^{2}}.$ Thus, the Larmor radius $%
\rho_0$ is constant, and the time dependence is contained in the uniform
gyration motion. Moreover, the time variation of the potential is slow. Its
characteristic time $\tau _{c}$ is very large compared to the gyration
period $\theta \equiv 2\pi /\Omega ,$ $\tau _{c}\gg \theta .$\ Since the
displacement of the guiding center during $\theta $\ is small, Eq. (\ref{csi}%
) can be averaged over the cyclotron period at constant $\boldsymbol{\xi }$\
and $t.$ One obtains%
\begin{equation}
\frac{d\xi _{i}}{dt}=-\varepsilon _{ij}\frac{\partial _{j}\psi (\boldsymbol{%
\xi },z,t;\rho_0)}{\partial \xi _{j}},  \label{ecgiro}
\end{equation}%
\begin{equation}
\psi (\boldsymbol{\xi },z,t;\rho_0)=\frac{1}{B}\frac{1}{\theta }%
\int_{t}^{t+\theta }d\tau \phi (\boldsymbol{\xi }(t)+\mathbf{\rho }(\tau
),z(t),t),  \label{figiro}
\end{equation}%
where the gyro-averaged potential $\psi $ was normalized with $B.$ Thus, the
motion of the guiding centers of the fast ions obeys the same equation as in
the limit of zero Larmor radius, but with the modified potential\ (\ref%
{figiro}). Using the Fourier representation of the potential, $\widetilde{%
\phi }(\boldsymbol{k},z(t),t),$ and performing the time integral 
\begin{equation}
\psi (\boldsymbol{\xi },z,t;\rho_0)=\frac{1}{B}\int dk_{1}dk_{2}%
\widetilde{\phi }(\boldsymbol{k},z(t),t)J_{0}(k\rho_0)\exp \left[ i%
\boldsymbol{k\cdot \xi }(t)\right] ,  \label{psi}
\end{equation}%
where the wave number $\boldsymbol{k=}\left( k_{1},~k_{2}\right) $ is
perpendicular on $\mathbf{B,}$ $k=\sqrt{k_{1}^{2}+k_{2}^{2}}$ and $J_{0}$\
is the Bessel function of the first kind.\ This shows that the gyro-average
of the potential $\phi $ determines the multiplication of its Fourier
transform with $J_{0}(k\rho_0),$ which corresponds to the gradual
attenuation of the large wave number components of the spectrum as the
Larmor radius increases.\ \ 

%%%%%%%%%%%%%%%%%%

%%%%%%%%%%%%%%%

\begin{figure}[h]
\subfloat{\includegraphics[width = 3in]{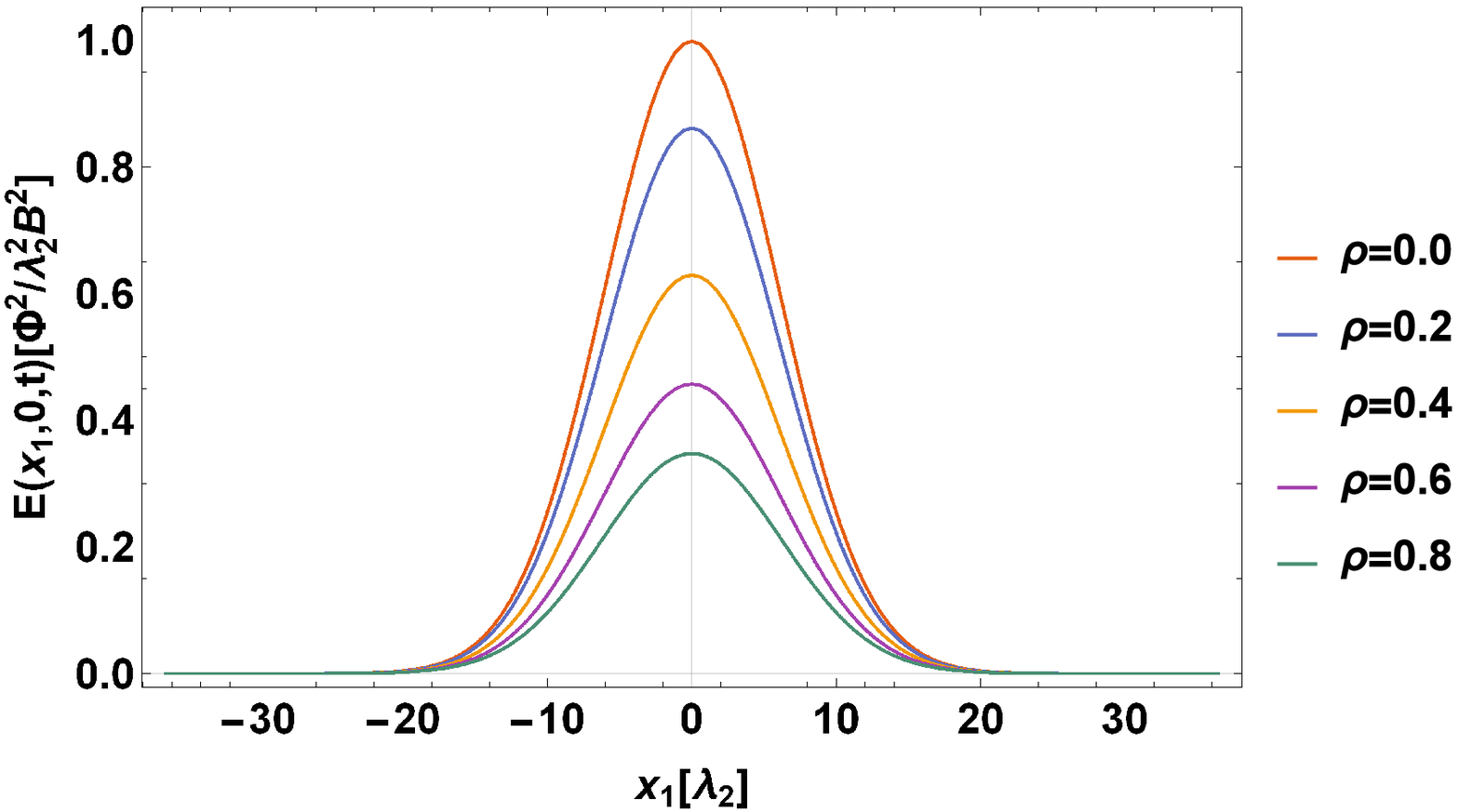}}
 \subfloat{%
\includegraphics[width = 3in]{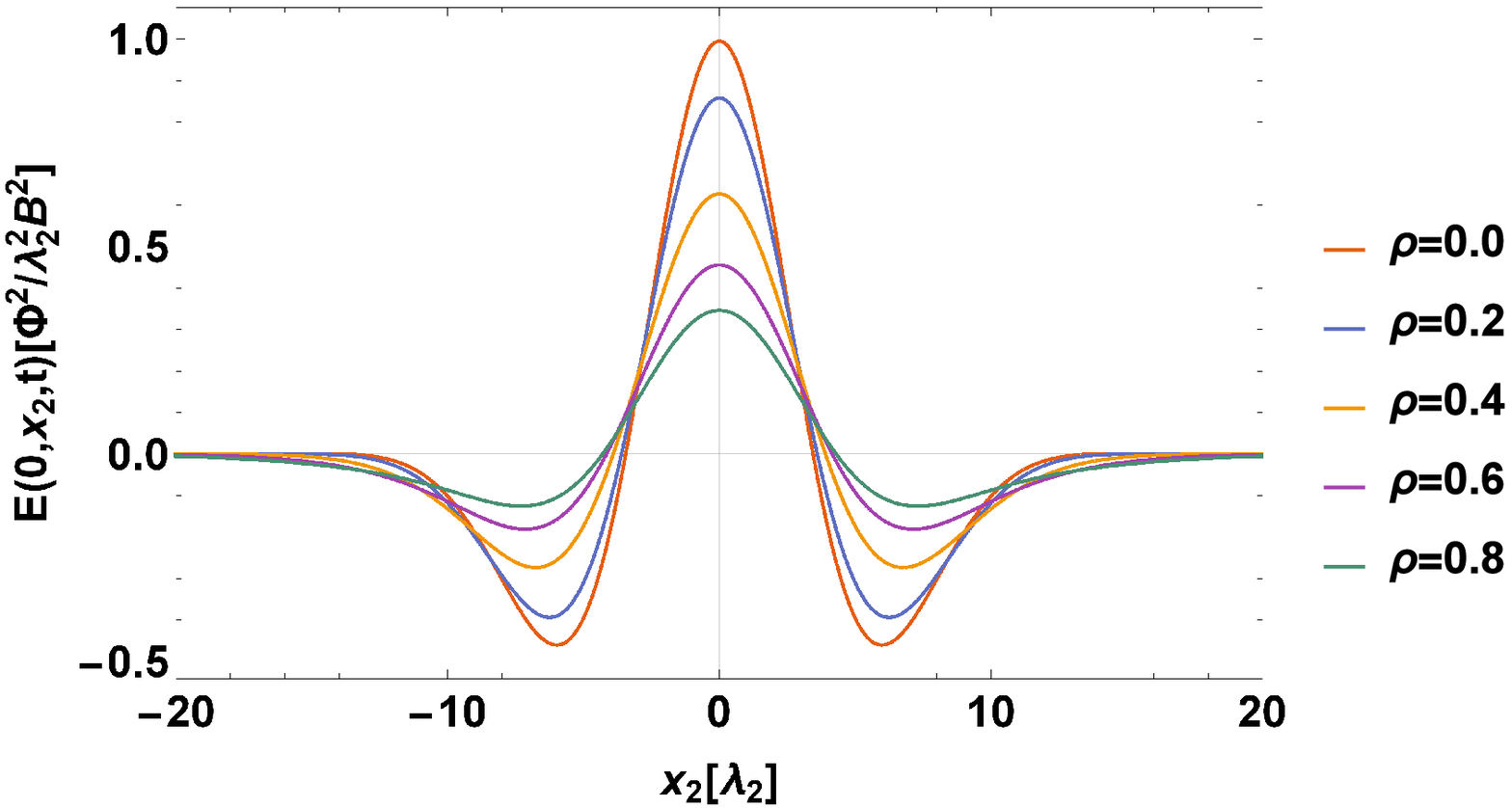}}
\caption{The EC (\ref{eff}) of the averaged potential $\protect\psi $ for the values of
the Larmor radius that label the curves.}
\label{fig1}
\end{figure}

\begin{figure}[tbp]
\centering
\includegraphics[width=0.7\linewidth]{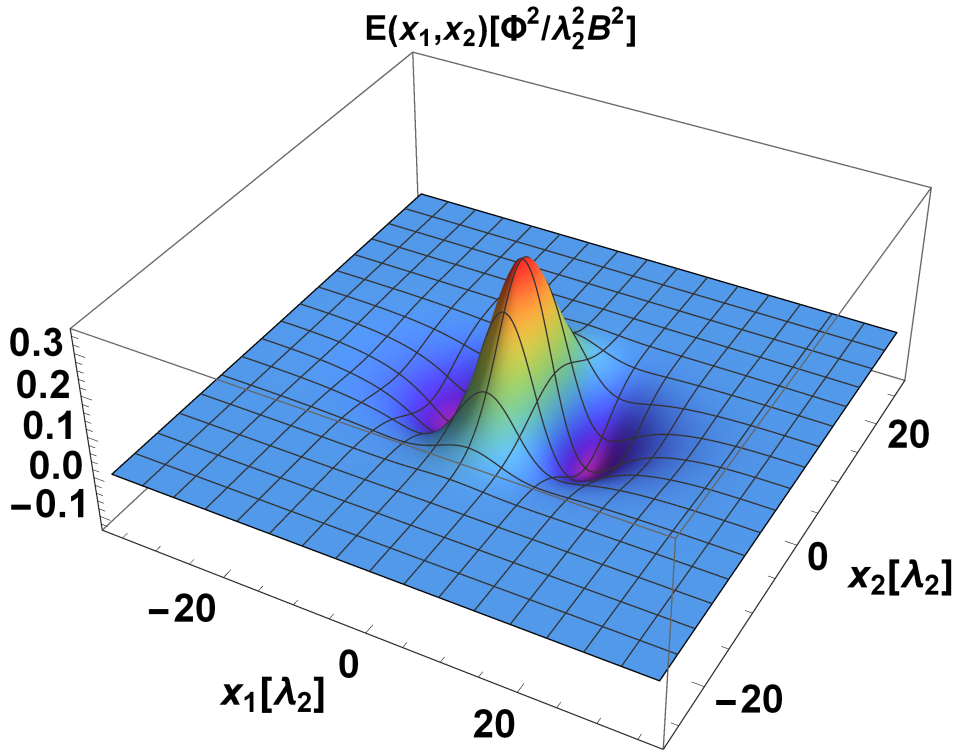}
\caption{$E(x_1,x_2,0,0;1.5)$ for $k_{0}=1$ and $a=9$.}
\label{fig2}
\end{figure}

The EC of the averaged potential $\psi (\boldsymbol{\xi },z,t;\rho _{\alpha
})$ for a Maxwellian distribution of particle velocities is \ 

\begin{eqnarray}
E(\boldsymbol{\xi },z,t;\rho _{\alpha }) &\equiv &\langle \psi (\boldsymbol{%
\xi }\mathbf{^{\prime }},z^{\prime },t^{\prime })\psi (\boldsymbol{\xi }%
\mathbf{^{\prime }}+\boldsymbol{\xi },z^{\prime }+z,t^{\prime }+t)\rangle 
\notag \\
&=&\frac{1}{B^2}\int d\mathbf{k}~S(\mathbf{k},z,t)\exp \left[ i\boldsymbol{k\cdot \xi 
}\right] \int_{0}^{\infty }dvv^{2}e^{-v^{2}}J_{0}^{2}\left( \frac{%
v}{\Omega _{\alpha }}k\right)  \notag \\
&=&\frac{1}{B^2}\int d\mathbf{k}~S(\mathbf{k},z,t)\exp \left( -\rho _{\alpha
}^{2}k^{2}\right) I_{0}(\rho _{\alpha }^{2}k^{2})~\exp \left[ i\boldsymbol{%
k\cdot \xi }\right] ,  \label{eff}
\end{eqnarray}%
\ \ where the velocity is given in units of the thermal velocity $v_{th}=\sqrt{T_{\alpha }/m_{\alpha }}$ of
the fast particles with the temperature $T_{\alpha },$ $\rho _{\alpha }=$\ $%
v_{th}/\Omega _{\alpha }$ and $I_{0}(x)$ the modified Bessel function of the
first kind. The limit $\rho _{\alpha }=0$ corresponds to the EC of the
potential $\phi .$ The EC (\ref{eff}) is represented in Figure \ref{fig1}
for several values of $\rho _{\alpha }.$ One can see that the amplitude of $%
\psi (\boldsymbol{\xi },z,t;\rho _{\alpha })$ is a monotonically decreasing
function of $\rho _{\alpha }$ and that the effective correlation lengths in the perpendicular plane increase with $\rho_\alpha$. The general shape of the EC is not changed
at large $\rho _{\alpha }.$ In particular, the positive and the negative
parts compensate, and the integral over $x_{2}$\ is zero for any $\rho
_{\alpha }.$\ This property is due to the spectrum (\ref{spectr}),\ which
cancels for $k_{2}=0.$ We plot in Figure \ref{fig2} the EC of the
gyro-averaged potential $\psi $ for $\bar{\rho }=\rho_\alpha/\lambda_2=1.5$, $k_{0}=1,$ $a=9$
and $\Phi =1.$

%%%%%%%%%%%%%%%%%%%%%

%%%%%%%%%%%%%%%%%%%%

\bigskip

%%%%%%%%%%%%%%%%%%%%

\subsection{The decorrelation trajectory method}

The transport of fast particles in turbulent plasmas is described by
equations that are similar to those for the $\mathbf{E}\times \mathbf{B}$\
drift. Moreover, the shape of the EC of $\psi $ is similar to that of $\phi
. $ We introduce normalized quantities \cite{vlad2013test} using as units $%
\lambda _{2}$ for the perpendicular displacements, $\lambda _{z}$ for parallel displacements, $%
V_{0}=\Phi /B\lambda _{2}$ for the $\mathbf{E}\times \mathbf{B}$ drift velocity and $%
\tau _{0}=\lambda _{2}/V_{0}$\ for time. The equations of motion for the
normalized quantities (designed by the same symbols as the physical ones) are%
\begin{equation}
\frac{d\xi _{i}}{dt}=-\varepsilon _{ij}\frac{\partial _{j}\psi (\boldsymbol{%
\xi },z,t;\bar{\rho})}{\partial \xi _{j}}+\delta _{i2}V_{d},~\ \ \frac{dz}{dt%
}=\frac{\tau _{0}}{\tau _{z}},\ \   \label{ecm}
\end{equation}%
where $\tau _{z}=\lambda _{z}/v_{th}$ is the decorrelation time induced by the
parallel motion, $\bar{\rho}=\rho _{\alpha
}/\lambda _{2}$, and $V_{d}=V_{\ast }/V_{0}$, with $V_\ast$ being the effective diamagnetic velocity. The equation is written in a frame that moves with the
potential, in which the normalized effective diamagnetic velocity appears as an average
velocity.

This type of statistical problem was analyzed in several papers. The time
dependent diffusion coefficients $D_{i}(t),~i=1,2$\ were determined using
the decorrelation trajectory method (DTM) \cite{vlad1998diffusion,vlad2004trajectory}. This
is a semi-analytical method, which shows that $D_{i}(t)$\ can be
approximated using a set trajectories obtained from the EC of the potential,
the decorrelation trajectories (DTs). The main idea of this method\ is to
group together trajectories that are similar by imposing supplementary
initial conditions. Each group corresponds to a subensemble $S$ of
realizations of the stochastic potential that is defined by the supplementary
initial conditions. One obtains a DT for each subensemble, which then is
used\ to evaluate $D_{i}(t)$ as weighted sums of the contributions of all
subensembles.

We use here the fast DTM introduced in \cite{vlad2015electron}, which imposes
only two supplementary initial conditions: the potential in the starting
point of the guiding centers trajectories $\psi ^{0}\equiv \psi (0,0,0;\bar{%
\rho})$\ and the orientation $\theta ^{0}$\ of the normalized initial
velocity. The advantage of this method is that the number of DTs is
strongly reduced.

The time dependent diffusion coefficients $D_{i}(t)$\ are obtained from%
\begin{eqnarray}
D_{1}(t) &=&\frac{V_{1}}{4}\int_{-\infty }^{\infty }d\phi ^{0}\exp \left( -%
\frac{\left( \phi ^{0}\right) ^{2}}{2}\right) \int_{0}^{2\pi }\ d\theta
^{0}\cos (\theta ^{0})X_{1}^{S}(t),  \label{Dprim} \\
D_{2}(t) &=&\frac{V_{2}}{4}\int_{-\infty }^{\infty }d\phi ^{0}\exp \left( -%
\frac{\left( \phi ^{0}\right) ^{2}}{2}\right) \int_{0}^{2\pi }\ d\theta
^{0}\sin (\theta ^{0})X_{2}^{S}(t),
\end{eqnarray}%
where $\mathbf{X}^{S}(t)$ is the DT in the subensemble $S$ that is the
solution of%
\begin{equation}
\frac{dX_{i}^{S}}{dt}=V_{i}^{S}(\mathbf{X}^{S},t)+\delta _{i2}V_{d}.
\label{DTprim}
\end{equation}%
The subensemble average velocities $V_{i}^{S}(\mathbf{x},t)$ are obtained
from the average potential%
\begin{equation}
\Phi ^{S}(\mathbf{x},t)=\phi ^{0}\frac{E(\mathbf{x},t;\overline{\rho })}{E(0)%
}+\sqrt{\frac{8}{\pi }}\cos (\theta ^{0})\frac{E_{2}(\mathbf{x},t;\overline{%
\rho })}{V_{1}}-\sqrt{\frac{8}{\pi }}\sin (\theta ^{0})\frac{E_{1}(\mathbf{x}%
,t;\overline{\rho })}{V_{2}},  \label{phiSprim}
\end{equation}%
\begin{equation}
V_{i}^{S}(\mathbf{x},t)=-\varepsilon _{ij}\frac{\partial }{\partial x_{j}}%
\Phi ^{S}(\mathbf{x},t),  \label{VSprim}
\end{equation}%
where $E_{i}(\mathbf{x},t;\overline{\rho })\equiv \partial E(\mathbf{x},t;%
\overline{\rho })/\partial x_{i}$ is the space derivative. An example of the
subensemble potential $\Phi ^{S}(\mathbf{x},0)$ is show in Figure \ref%
{fig3}. The transport at large space and time scales is described by
the asymptotic value $D_{i}^{\infty }\equiv \underset{t\rightarrow \infty }{%
\lim }D_{i}(t).$

A computer code was developed for determining $D_{i}(t)$\ using Eqs. (\ref%
{Dprim}-\ref{VSprim}). It calculates the EC of $\psi $\ (\ref{eff}) and its
derivatives that appear in the subensemble average velocity (\ref{VSprim})\
using a fast Fourier Transform subroutine. The latter links an uniform grid representation in the ${\bf k}$ space to a two-dimensional real space mesh on which the velocity field is computed. The spectrum (\ref{spectr}) is used for all the
calculations presented in this paper, but it can easily be replaced by other
models. The DTs are determined using high order interpolation techniques for the velocity field. The time step automatically adapts so that
only the space steps along $x_{1}$ and $x_{2}$\ have to be optimized. The
condition is provided by the trajectories with $\tau _{d}\rightarrow \infty
, $\ which represent periodic motions on the contour lines of $\psi .$ They
have to remain close to these lines during the whole integration time that
can be of hundreds of periods.

\ \ %%%%%%%%%%%%%%%%%

\begin{figure}[tbp]
\centering
\includegraphics[width=0.7\linewidth]{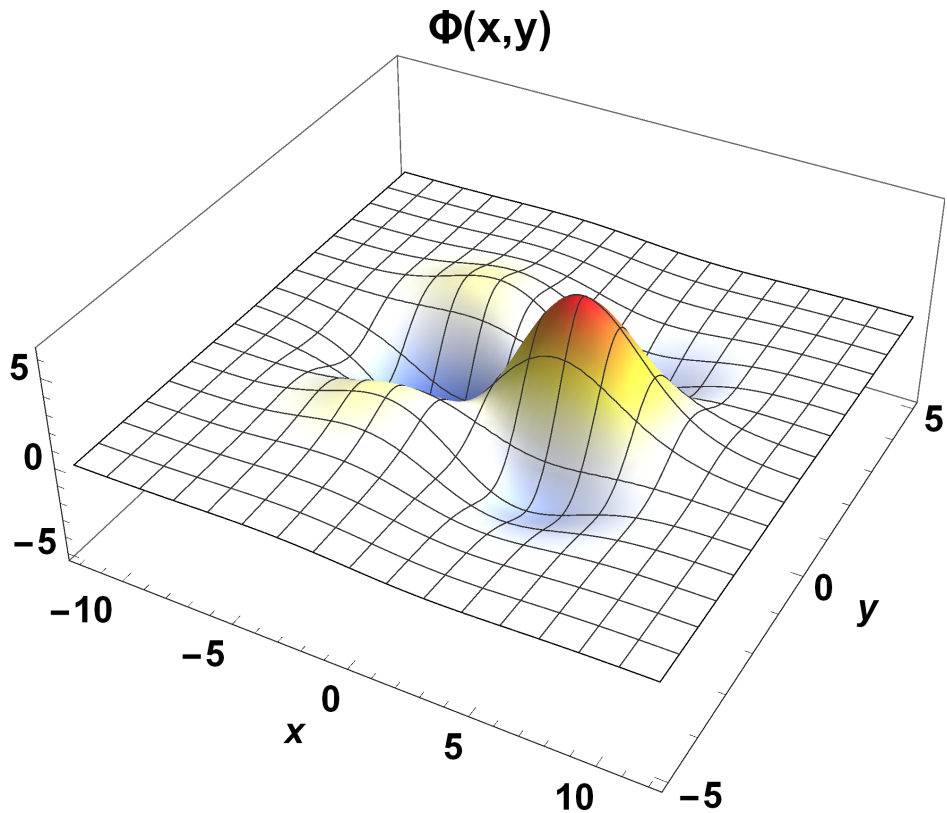}
\caption{The subensemble average potential $\Phi(x_1,x_2)$ defined by Eq. 
\ref{phiSprim} for $\protect\bar{\rho}=1.5$, $k_{0}=1$, $a=9$, $\protect\phi%
^0=1$, $v_1^0=2$, $v_2^0=3$ and $V_d=0$.}
\label{fig3}
\end{figure}

%%%%%%%%%%%%%%%%%

\section{Fast particle diffusion regimes}\label{results}

\subsection{Basic physical processes}

The diffusion regimes of the electrons \cite{vlad2015electron} and of the ions \cite{vlad2013test} in
the realistic model of the spectrum (\ref{spectr}) were studied for the
limit of zero Larmor radius.
The EC of the gyro-average potential (\ref{eff}) contains eight physical
parameters, six from the spectrum of the turbulence (\ref{spectr}): the amplitude of the potential fluctuations, $\Phi$, the maximum wave number $k_0$, the spatial decorrelation lengths $\lambda_1, \lambda_2, \lambda_z$ and the temporal decorrelation length $\tau_c$ plus the
Larmor radius, and the effective diamagnetic velocity $%
V_{\ast }$. The latter appears as the drift of the potential in the poloidal
direction.
 We have found that the transport regimes are
determined by four dimensionless parameters $K_{d},$ $K_{\ast }$ or equivalently $V_{d},$ 
$k_{0}$\ and\ $a,$\ which are defined by%
\begin{eqnarray}
K_{d} &\equiv &\frac{\tau _{d}}{\tau _{0}},~\ \tau _{d}=\frac{\tau _{c}\tau
_{z}}{\tau _{c}+\tau _{z}},~  \label{Kd} \\
K_{\ast } &\equiv &\frac{\tau _{\ast }}{\tau _{fly}}=\frac{V_{y}}{V_{d}},\ \
V_{d}=\frac{V_{\ast }}{V_{0}},  \label{Kst} \\
k_{0} &\equiv &k_{0}\lambda _{2},  \label{k0} \\
a &=&\frac{\lambda _{1}^{2}}{\lambda _{2}^{2}}.  \label{a}
\end{eqnarray}%
\ The values of $K_{d}$\ and $K_{\ast },$ are related to the presence
of trajectory trapping or eddying in the structure of the stochastic
potential.
% determine the transport regimes.

The effective Kubo number $K_{d}$ (\ref{Kd}) is a measure of the
decorrelation of the trajectories from the potential, which is determined by
the time variation of the potential and/or by the parallel motion of the
particles. Trajectory trapping exists when the decorrelation is weak such
that the characteristic time $\tau _{d}$ is larger than the time of flight $%
\tau _{fl}=(\lambda _{x}/V_{x}+\lambda _{y}/V_{y}).$\ 

The diamagnetic parameter $K_{\ast }$ (\ref{Kst}) is a dimensionless measure
of the effective velocity $V_{d},$\ which determines the characteristic time 
$\tau _{\ast }=\lambda _{y}/V_{d}.$ It is equivalent with an average
potential $xV_{d}$, which adds to the stochastic potential and changes the
configuration of the total potential. Bunches of opened contour lines appear
on a fraction of the surface that increases from zero (for $V_{d}=0,$ $%
K_{\ast }=\infty )$\ to one (for $V_{d}>V,~\ K_{\ast }<1).$\ Trajectory
trapping is possible only when the bunches of open lines fill only a
fraction of the surface, and islands of closed contour lines exists between
them. This configuration corresponds to the condition $V_{d}<V,~$ $K_{\ast
}>1.$ \ 

The main wave number $k_{0}$\ (\ref{k0}) and the anisotropy\ $a$\ (\ref{a})
influence the shape of the contour lines of the potential, and they only
lead to changes of the parameters of the transport regimes.

The special shape of the spectrum (\ref{spectr}) and of the EC (Figure \ref{fig1})
leads to similar dependences of $D_{1}^{\infty }$ on $K_{d}$ for the
quasilinear regime ($K_{\ast }<1)$\ and for the nonlinear regime ($K_{\ast
}>1)$ \cite{vlad2015electron}. In both cases, $D_{1}^{\infty }=V_{1}^{2}\tau _{d}$ for
small $K_{d},$\ then it has a maximum at $K_{d}=K_{\max }$\ and eventually
it decays as $K_{d}^{-\nu }.$\ $K_{\max }$ depends on the transport regime: $%
K_{\max }=K_{\ast }\sqrt{a}/\sqrt{k_{0}^{2}+3}$\ for $K_{\ast }<1,$ and $%
K_{\max }=\sqrt{a}/\sqrt{k_{0}^{2}+3}$\ for $K_{\ast }>1.$\ The power $\nu $
also depends on the regime and on the EC of the potential. We note that, for
an usual decreasing EC without negative minima, the diffusion coefficient in
the quasilinear regime ($K_{\ast }<1)$\ has significantly larger values at
large $K_{d}$\ since it saturates at the maximum value instead of decaying.\
\ \ \ 

\ \ \ \ \ \ \ \ \ \ \ 

%%%%%%%%%%%%%%%%%%%

\begin{figure}[h]
\subfloat{\includegraphics[width = 3.3in]{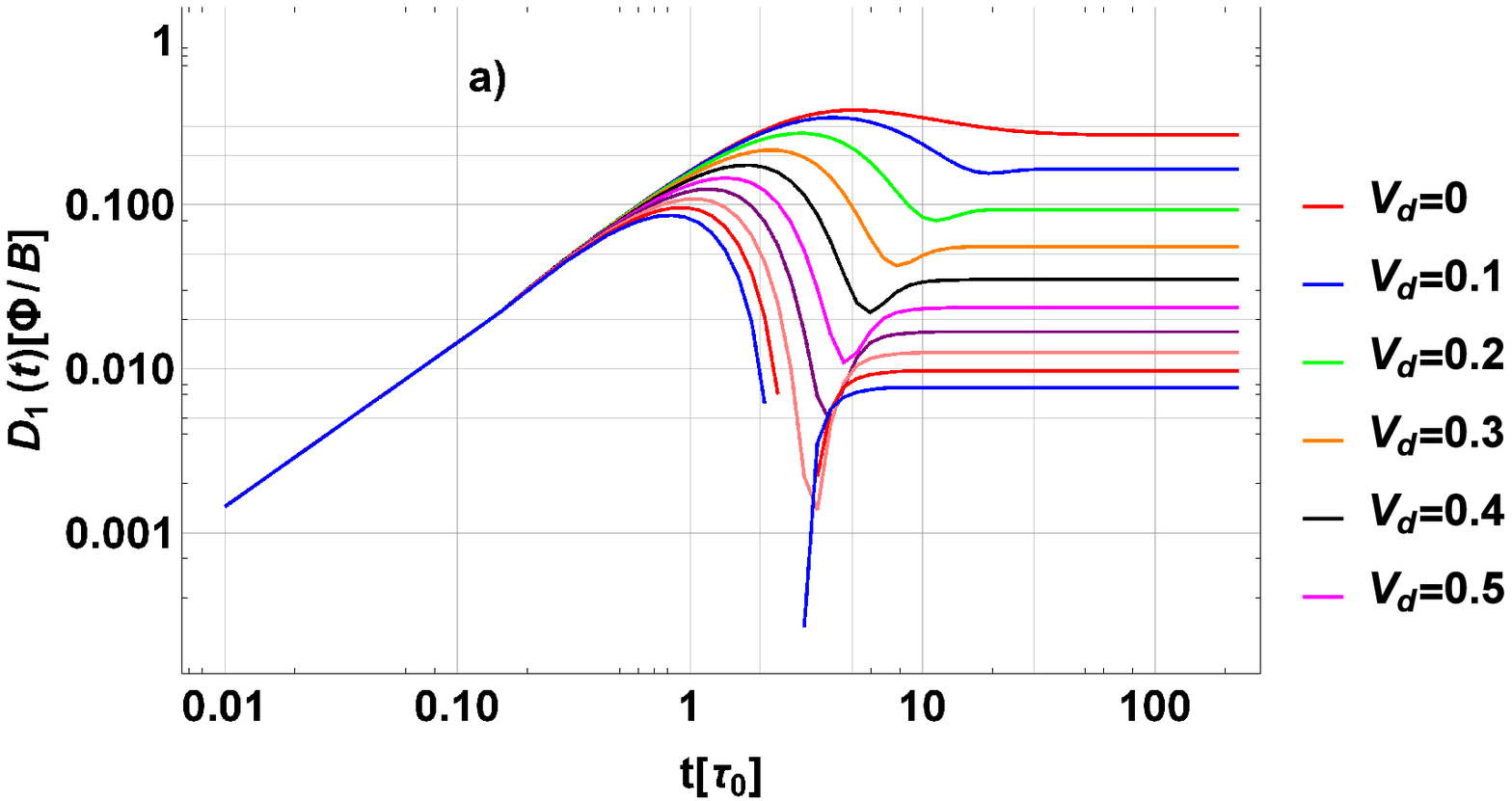}} \subfloat{%
\includegraphics[width = 3.3in]{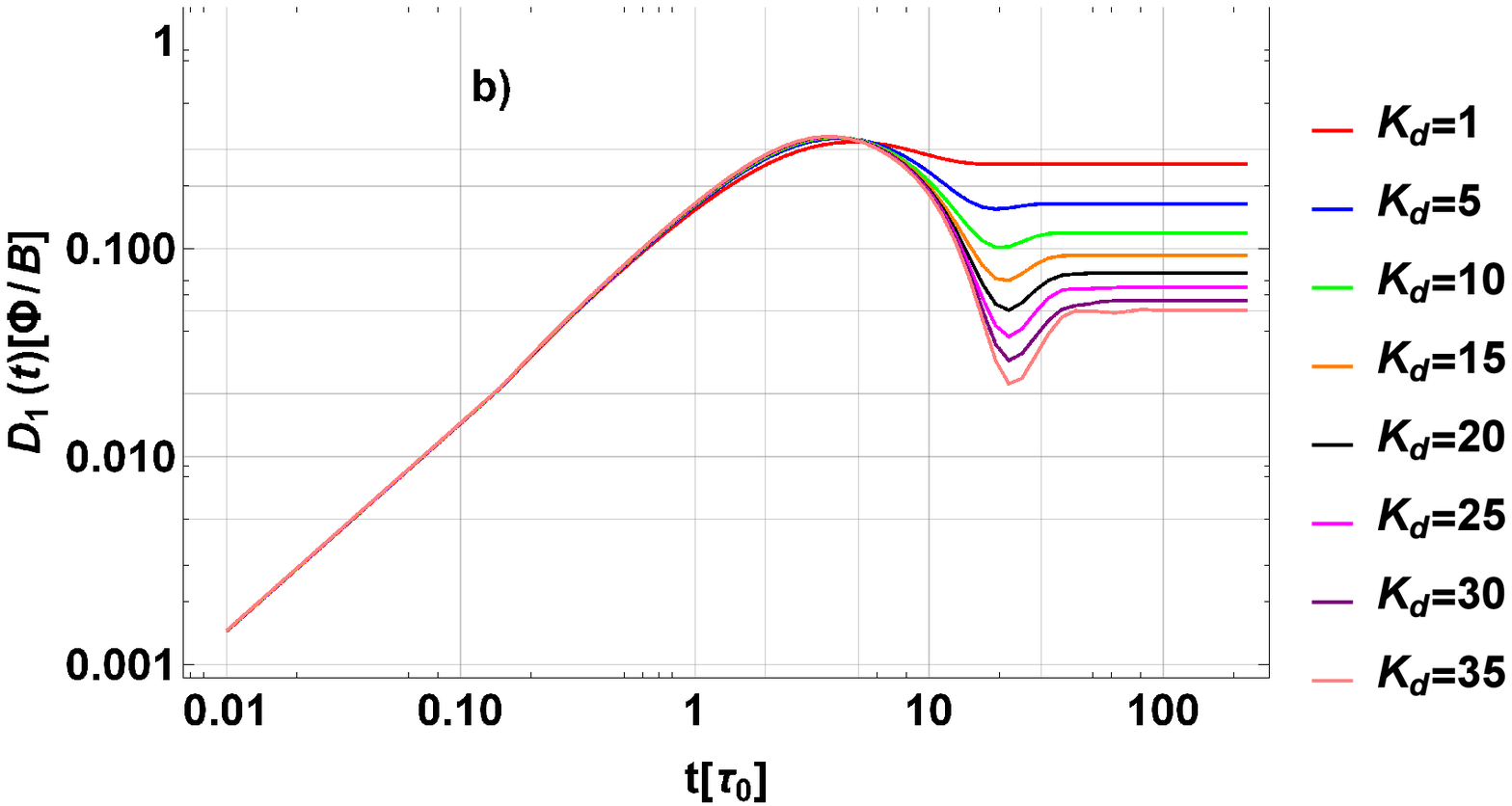}}
\caption{a) The time dependent diffusion coefficient $D_1(t)$ for several
values of $V_d$ and b) of $K_d$. The other parameters
are $\protect\bar{\rho}=4$ , $k_{0}=1$ and $a=9$}
\label{fig4}
\end{figure}

%%%%%%%%%%%%%%%%%%%

\begin{figure}[h]
\subfloat{\includegraphics[width = 3.3in]{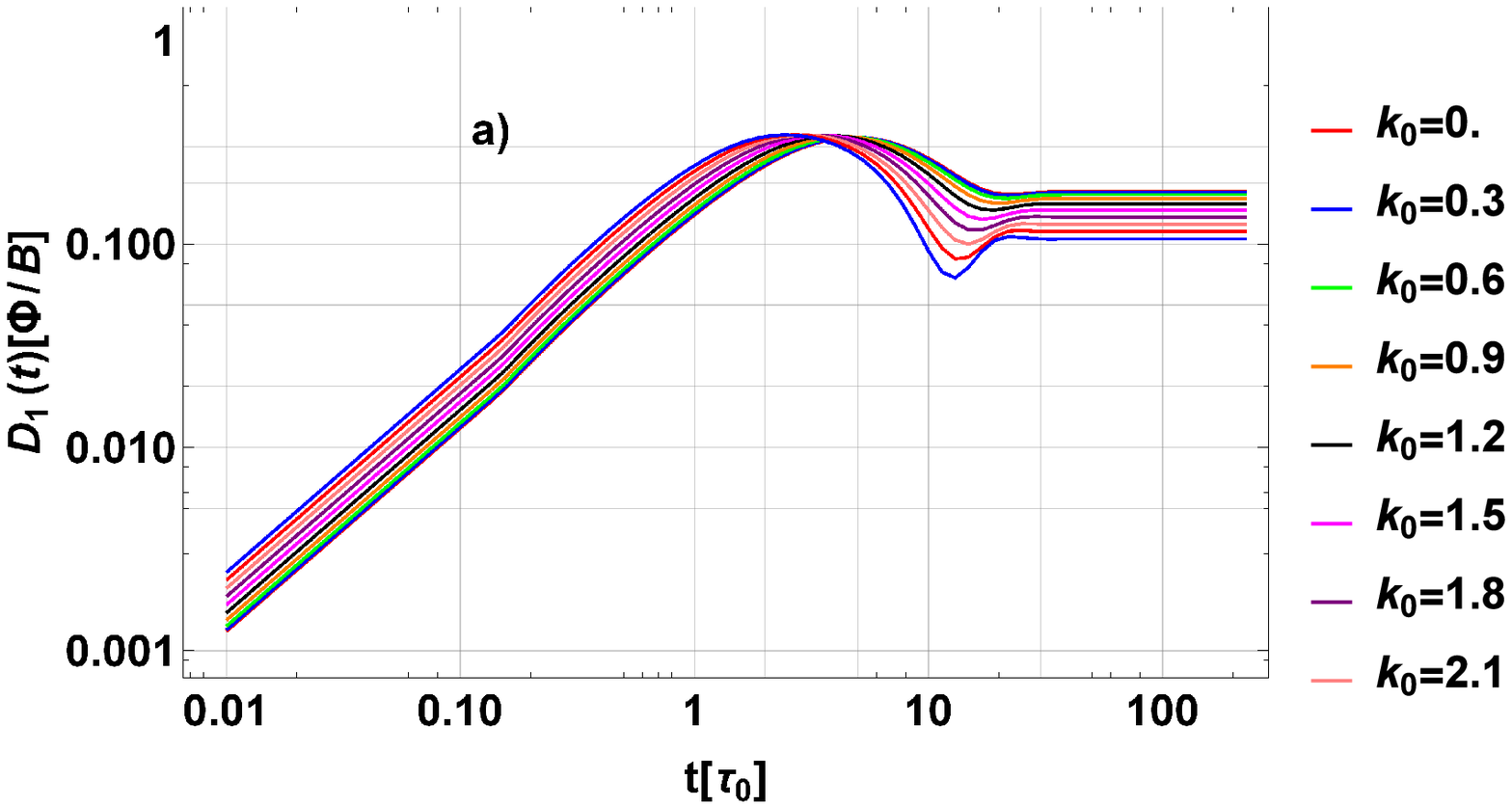}} \subfloat{%
\includegraphics[width = 3.3in]{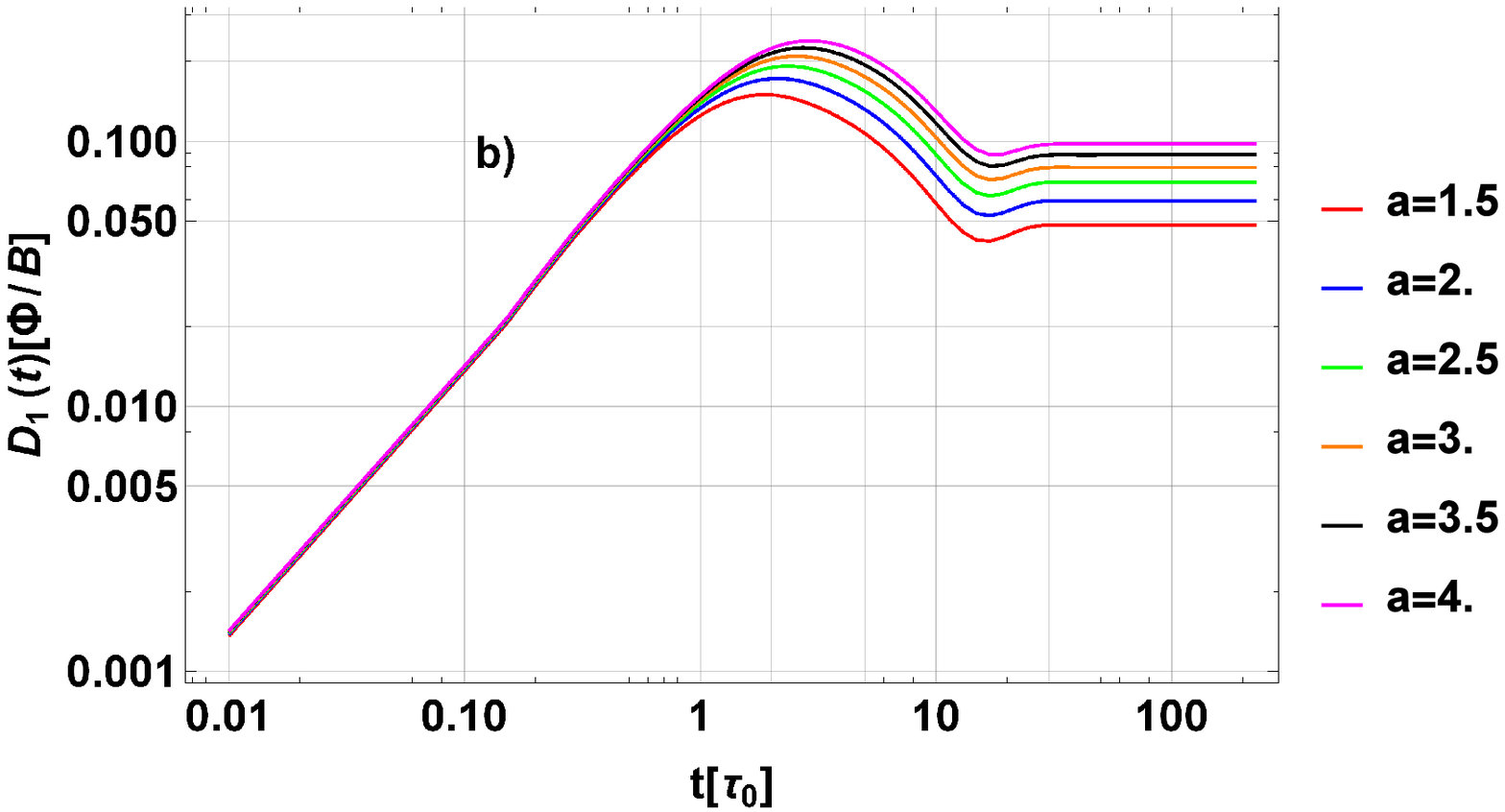}}
\caption{a) The time dependent diffusion coefficient $D_1(t)$ for several
values of $k_{0}$ and b) of $a$. The other parameters
are $\protect\bar{\rho}=4$, $K_d=10$ and $V_{d}=0.1$.}
\label{fig5}
\end{figure}

%%%%%%%%%%%%%%%%%

\begin{figure}[h]
\subfloat{\includegraphics[width = 3.3in]{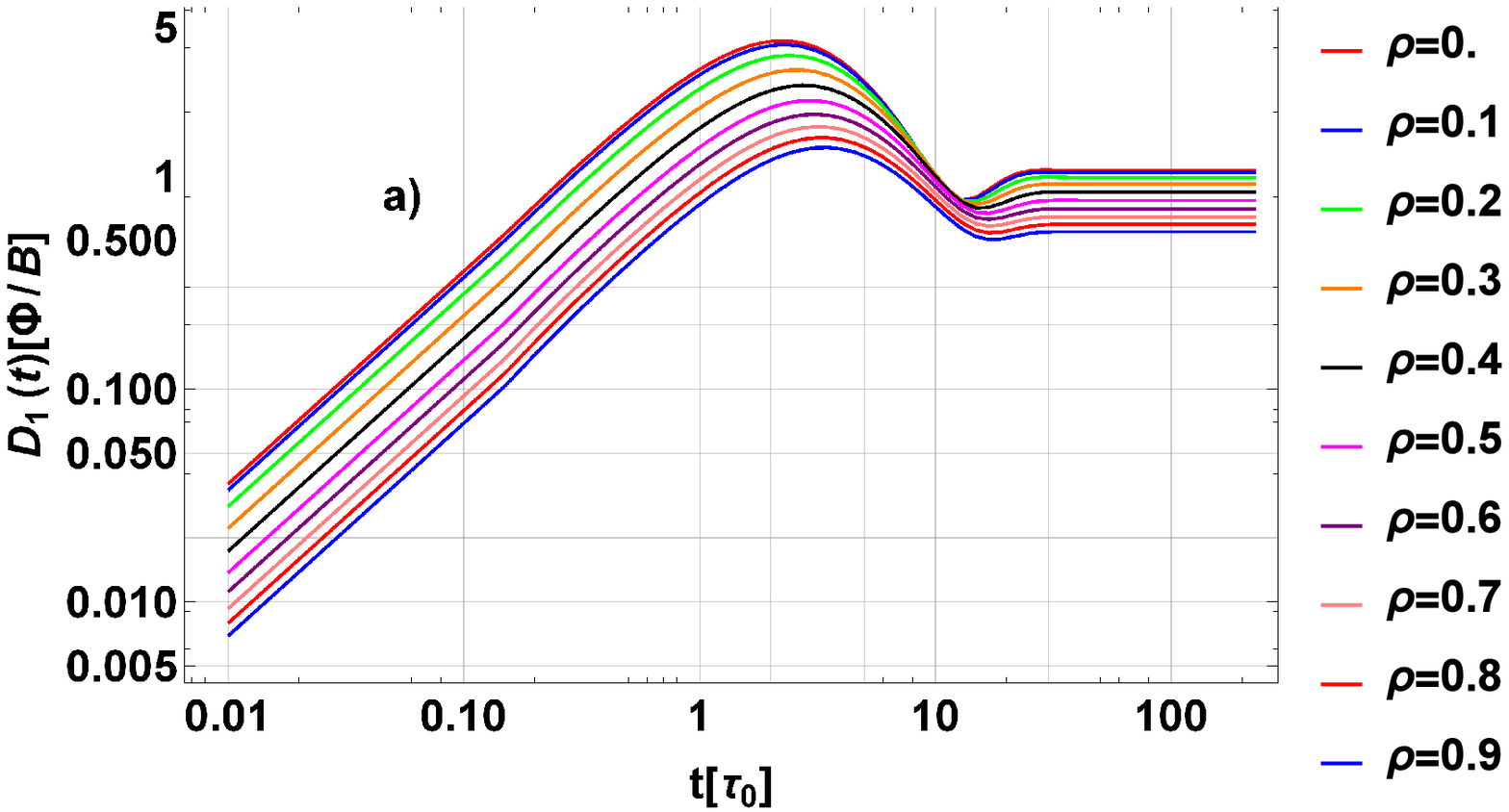}} \subfloat{%
\includegraphics[width = 3.3in]{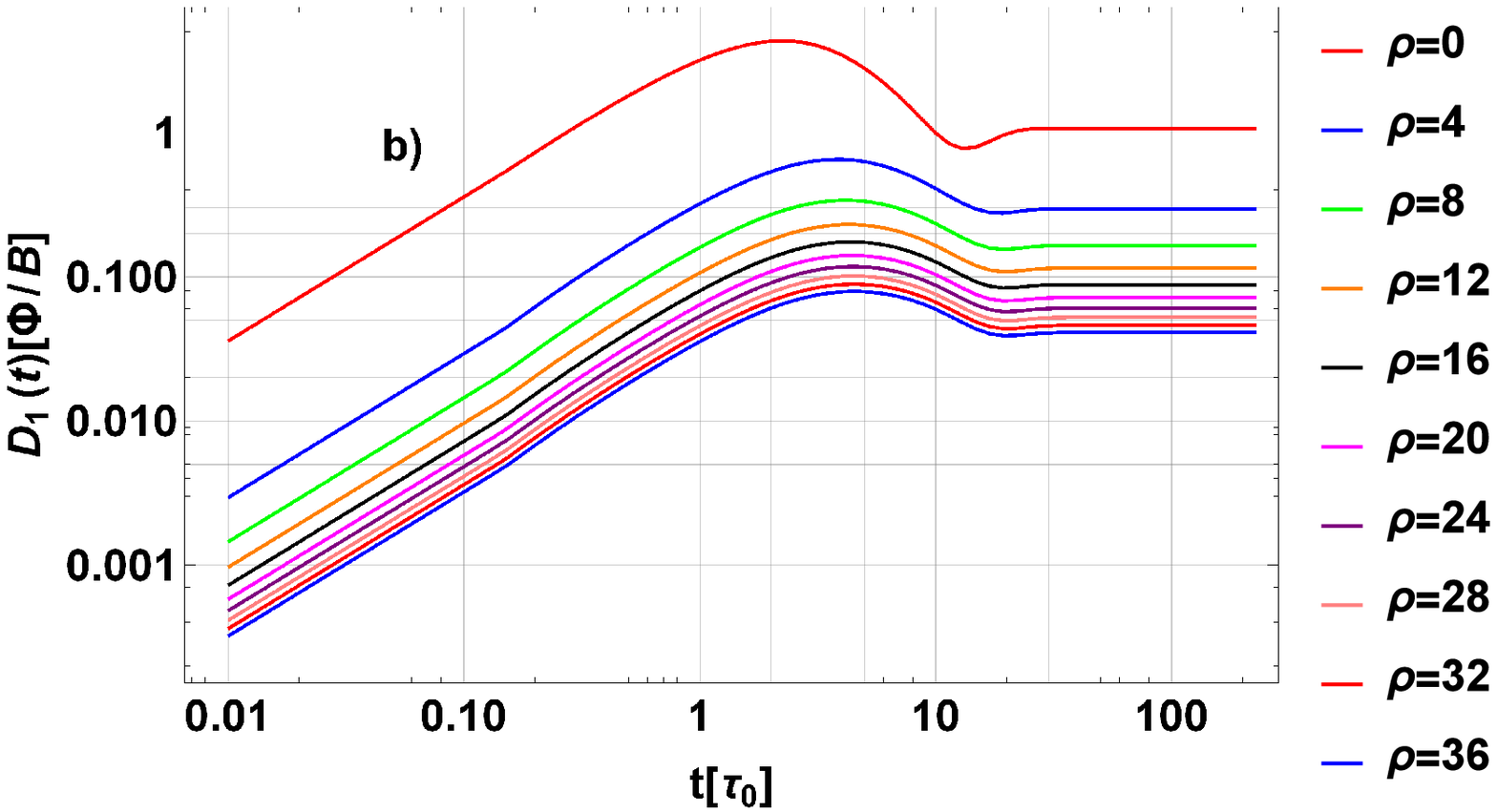}}\ 
\caption{a) The time dependent diffusion coefficient $D_1(t)$ for $\protect\bar{\rho}%
\in[0,1]$  and b) for $\protect\bar{\rho}\in[0,18]$ with $K_d=10$ and $V_{d}=0.1$ }
\label{fig6}
\end{figure}

%%%%%%%%%%%%%%%%%%

The first question addressed in this paper concerns the basic physics of the
transport regimes of the fast particles. We examine the dependence of the
diffusion coefficient on the four parameters at a large value of the Larmor
radius ($\overline{\rho }=4).$ Examples of the time dependent diffusion
coefficients obtained using the DTM in the nonlinear regime are shown in
Figures \ref{fig4} and \ref{fig5}.

The effects of the decorrelation and of the average velocity in the
nonlinear regime are presented in Figure \ref{fig4}. The decorrelation leads to the
transition from the subdiffusive to diffusive transport by saturating $%
D_{i}(t).$ The process is similar with the case of small energy particles.
The saturation time increases with the increase of $K_{d}$ and the
asymptotic diffusion coefficient decreases (Figure \ref{fig4}b). This is
due to the fraction of non-trapped trajectories that decreases when $K_{d}$\
increases, leading to the decrease of $D_{1}^{\infty }.$ The effect of the
average velocity for a large value of the decorrelation parameter ($%
K_{d}=10) $ (Figure \ref{fig4}a) consists of the continuous decrease of
the radial diffusion when $\overline{V}_{d}$\ increases.

The effect of the dominant wave number $k_{0}$\ is shown in Figure \ref{fig5}a. The increase of $k_{0}$\ determines the decrease of the radial
diffusion\ coefficient in the nonlinear regime, although it leads to the
increase of the amplitude of the radial velocity ($V_{1}=V_{0}\sqrt{%
k_{0}^{2}+3}$). \ \ \ \ 

Thus, at large Larmor radius, the time dependent diffusion coefficients are
qualitatively similar to those for $\overline{\rho }=0.$ This is also
suggested by the shape of the gyro-averaged EC (\ref{eff}), which is not much changed
compared to that for $\overline{\rho }=0.$ \ 

\subsection{Fast particle transport regimes}

The dependence of the diffusion coefficient $D_{1}(t)$ on $\overline{\rho }$%
\ is shown in Figure \ref{fig6}. The shapes of the curves are roughly similar for
different values of $\overline{\rho }.$ The differences between the curves
with different $\overline{\rho }$\ are practically independent on time,
except for the range $\overline{\rho }<1$\ where a stronger dependence on
time can be seen (Figure \ref{fig6}a).

\begin{figure}[h]
\subfloat{\includegraphics[width = 3.4in]{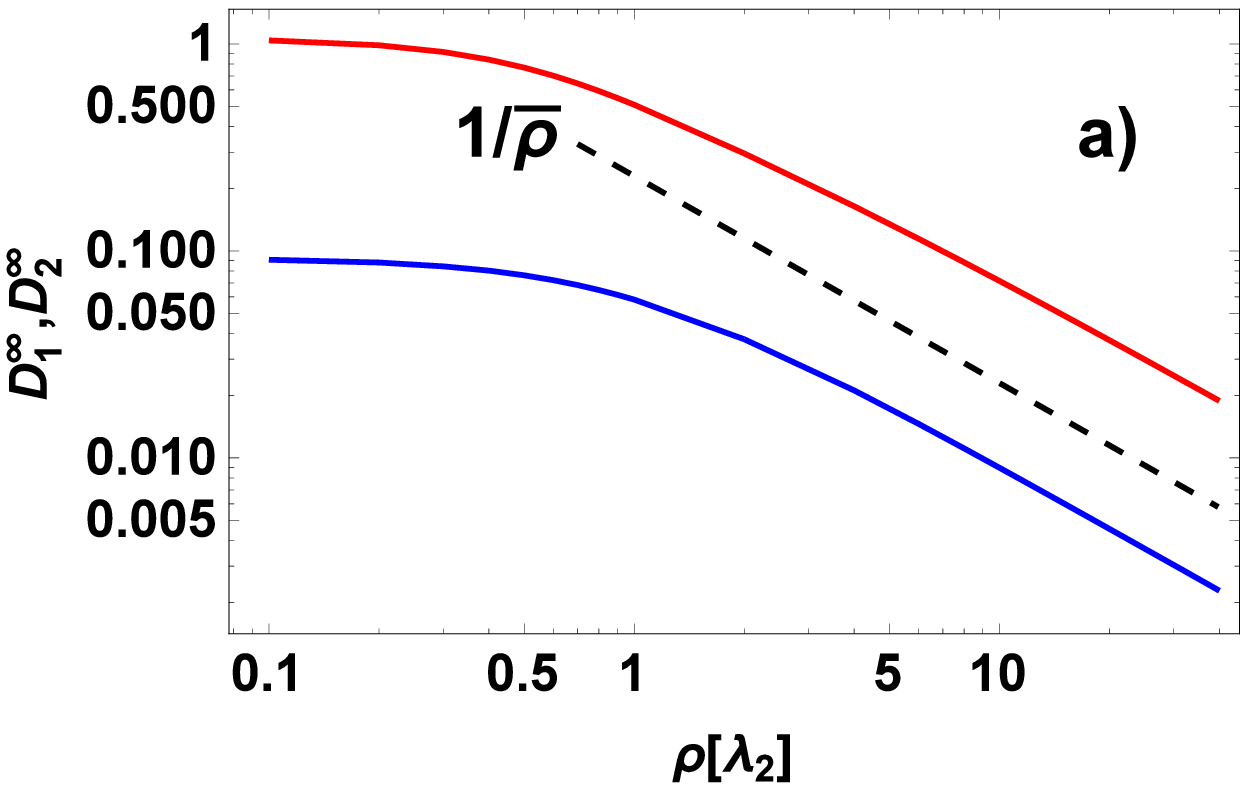}} \subfloat{%
\includegraphics[width = 3.4in]{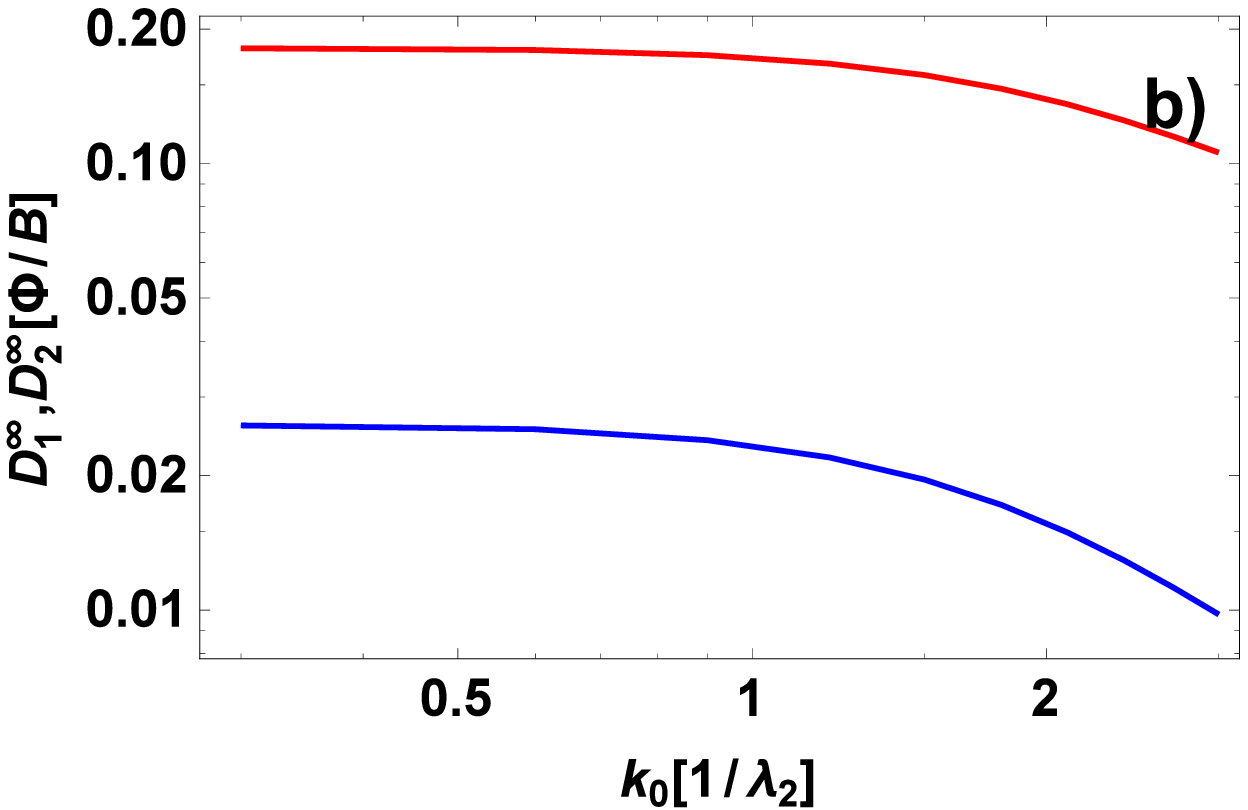}}\ \subfloat{%
\includegraphics[width = 3.4in]{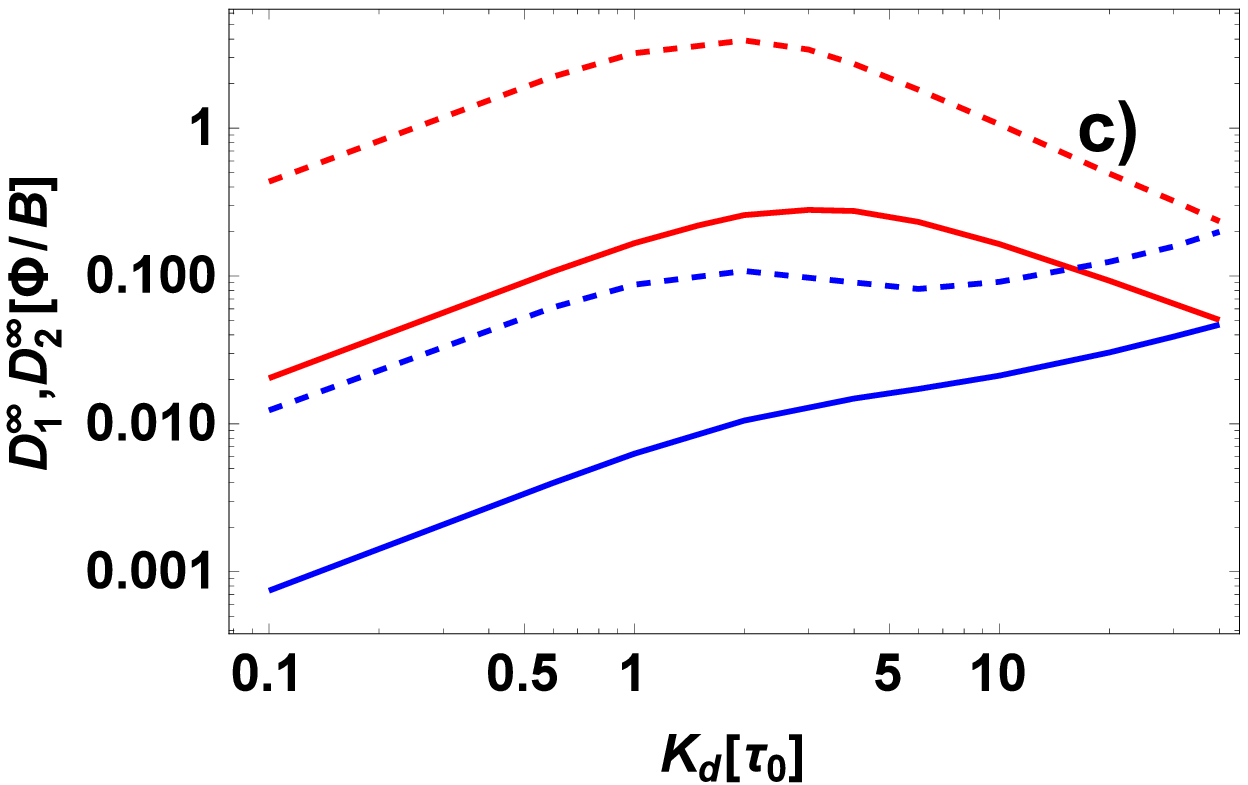}} \subfloat{%
\includegraphics[width = 3.4in]{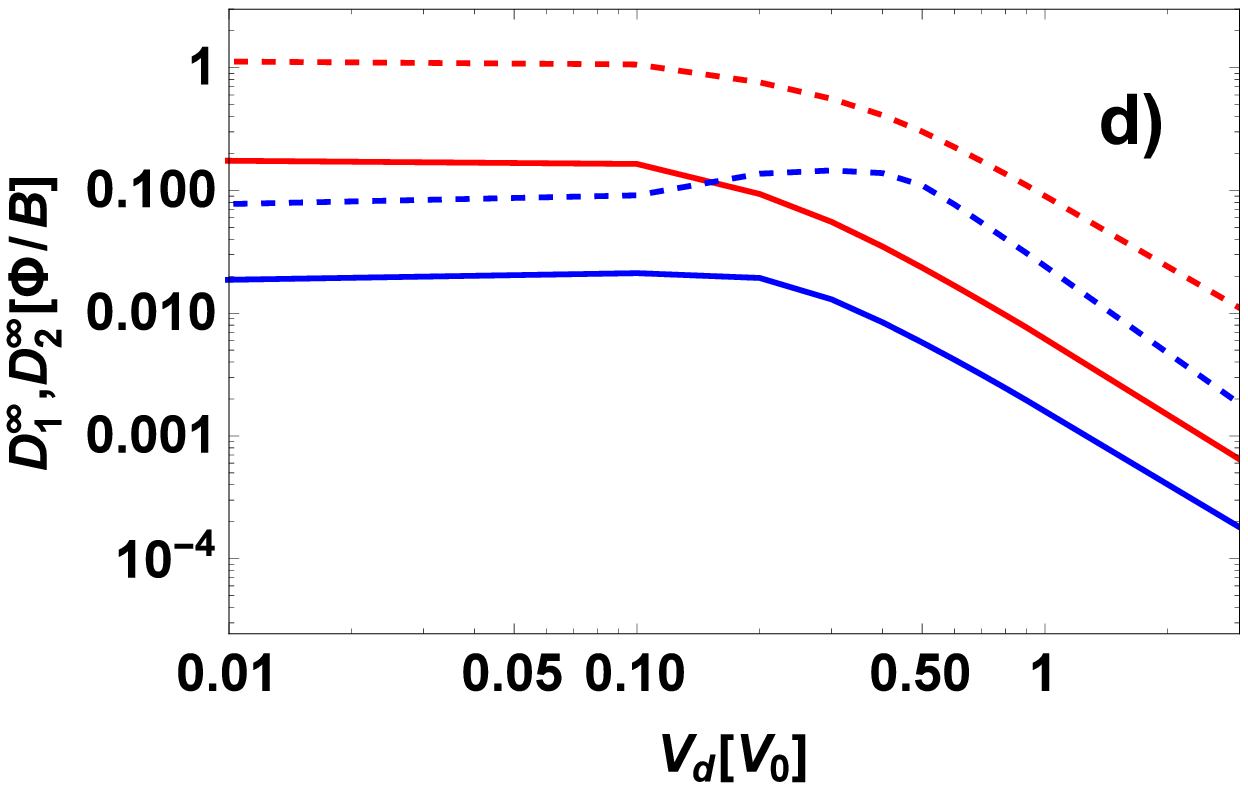}}
\caption{The asymptotic diffusion coefficients $D_1^\infty$ (red line) and $%
D_2^\infty$ (blue line). The dashed colored lines correspond to $\protect\bar{\rho}=0$.
The constant parameters are $\bar{\rho}=4,  k_0=1, a=9, K_d=10$ and $V_{d}=0.1$ . The dashed line in a) corresponds to the function $1/\bar{\rho}$. }
\label{fig7}
\end{figure}

The asymptotic diffusion coefficients $D_{i}^{\infty }$ are represented in
Figure \ref{fig7} in the nonlinear regime. Figure \ref{fig7}a shows the dependence
on the Larmor radius $\overline{\rho }.$\ One can see that both $%
D_{1}^{\infty }$\ and $D_{2}^{\infty }$\ do not depend on $\overline{\rho }$
for $\overline{\rho }<<1,$ and that they decay at large $\overline{\rho }$\
\ as $1/\overline{\rho }.$\ This decay is the same as in the quasilinear
regime. Fast particle diffusion coefficient can be evaluated analytically in
the case of the quasilinear (Gaussian) transport. It decreases as $1/%
\overline{\rho }.$\ The similar dependence on $\overline{\rho }$ in the
nonlinear and quasilinear regimes is rather surprising because several works
have found a weaker decay in the nonlinear regime (as $1/\overline{\rho }%
^{0.38}$ in \cite{hauff2007b}).

The cause of the faster decay with $\overline{\rho }$\ found here is the
special shape (\ref{spectr}) of the spectrum of the drift type turbulence.
The average over the gyro motion leads to the attenuation of the large $k$
components of the spectrum ($k\overline{\rho }\gtrsim 1)$, while the small $%
k $ ($k\overline{\rho }\ll 1)$ part of $S$ is not affected. The spectrum (%
\ref{spectr}) decays in the small $k$ range because the modes are stable for 
$k_{2}=0.$ Due to this property, all the components of the spectrum are
attenuated at large enough vales of $\overline{\rho }$ because the condition 
$k\overline{\rho }\gtrsim 1$\ applies for all components that correspond to
significant (not close to zero) values of $S.$ This leads at large values of $%
\overline{\rho }$ to the change of the effective EC that consists only in
the decay of the amplitude but not in the modification of the shape (increase
of the correlation lengths like for monotonically decaying spectra).

From Figure \ref{fig7}b one can see that the asymptotical value of the diffusion coefficients $D_{i}^{\infty }$  does not depend on the value of the $k_0$ wave vector for $k_0<1$ and that is has a weak exponential decay at large $k_0$.

The dependence of $D_{i}^{\infty }$ of the fast ions on the decorrelation
parameter $K_{d}$\ is shown in Figure \ref{fig7}c. One can see that the
diffusion coefficients are smaller than in the zero Larmor radius limit
(dashed curves).\ The dependence on $%
K_{d}$\ of the fast particle diffusion coefficient is the same as at $\rho
=0 $\ for both limits of small and large $K_{d}$\ (the curves are parallel
in these limits). The maximum of $D_{1}^{\infty }(K_{d})$ is displaced to
larger values of $K_{d}$\ at large $\overline{\rho }.$
In Figure\ref{fig7}d is plotted the dependence of $D_{i}^{\infty }$ on the normalized effective diamagnetic velocity $V_d$. The behaviour is similar as in the case of vanishing Larmor radius (dashed lines) both in the nonlinear and in the quasilinear regimes. 
\ \ 

%%%%%%%%%%%%%%%%%

%%%%%%%%%%%%%%%%%

%%%%%%%%%%%%%%%%%

\begin{figure}[h]
\subfloat{\includegraphics[width = 6in]{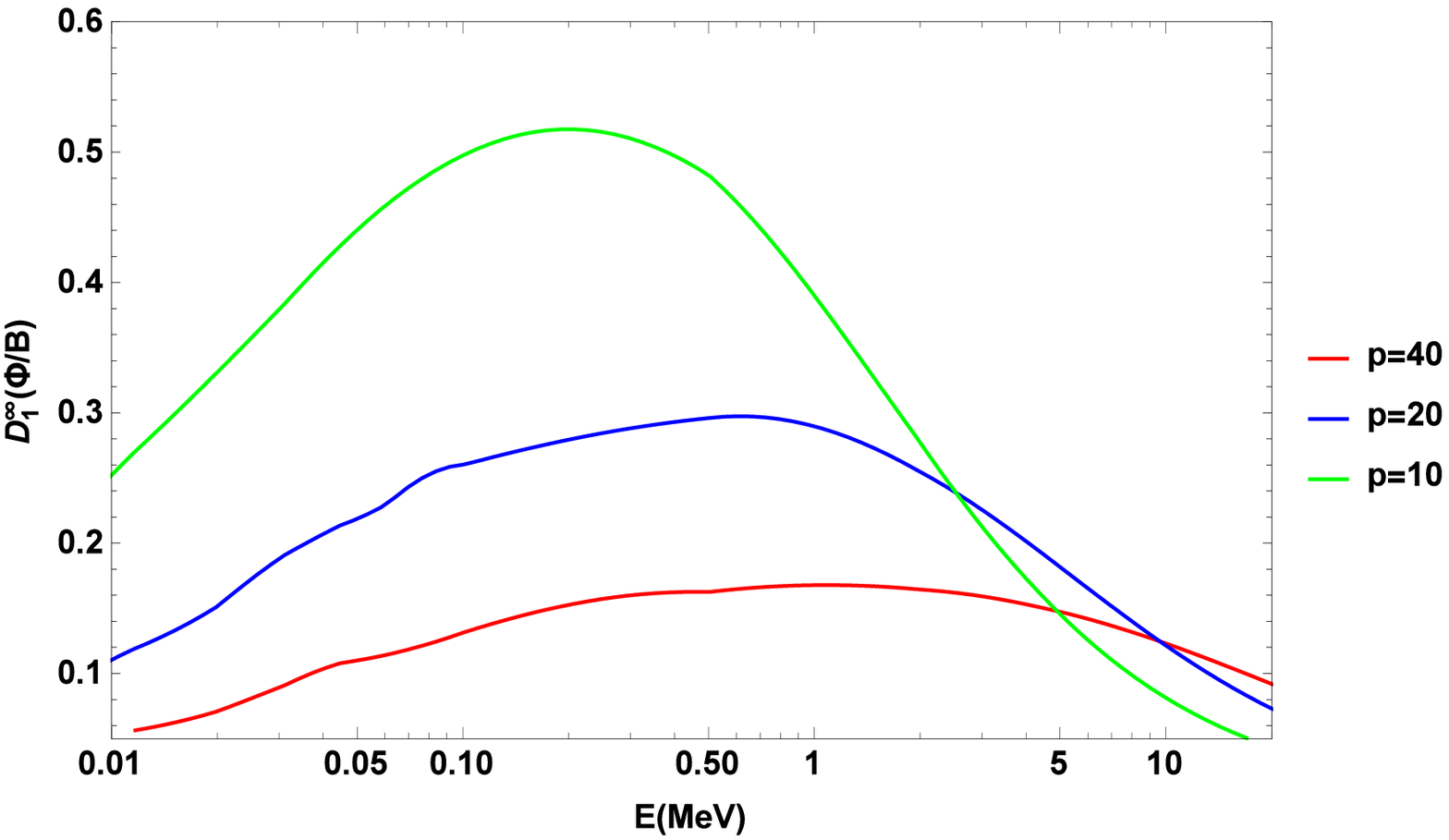}}
\caption{Asymptotic values of diffusion coefficient $D_1^\infty$ as function
of the energy for the values of $p$ that label the curves.}
\label{fig8}
\end{figure}

%%%%%%%%%%%%%%%%

\subsection{Alpha particle turbulent loss as function of the energy}

Numerical simulations have shown that the main decorrelation mechanism in
ITG turbulence is the parrallel ion motion. The turbulent transport of the
fast particles, which have much smaller parallel time $\tau _{z},$ is
completely dominated by the parallel decorrelation. The decorrelation
parameter (\ref{Kd}) becomes $K_{d}\cong \tau _{z}$ in these conditions.

Both the decorrelation time and the Larmor radius are functions of the
energy of the $\alpha $ particles. The cooling of the fast particles
determine the decrease of $\bar{\rho}$ and the increase of $\tau _{z},$%
\ but the product of these parameters $\bar{\rho}\tau _{z}=(m_{\alpha
}/q_{\alpha })(\lambda _{z}/B)$\ does not depend on $\alpha $ particle
energy. It essentially depends on plasma size through $\lambda _{z}.$\ In
terms of the dimensionless parameters that characterize fast particle
transport%
\begin{equation}
\bar{\rho}\tau_{z}=p,\ \ p\equiv \mathcal{V}%
\frac{\lambda _{z}}{R}\frac{R}{L_{T_{i}}}\left( \frac{\rho _{ion}}{\lambda _{y}%
}\right) ^{2},~\ \ \mathcal{V}\equiv \frac{\Phi }{B\lambda _{y}V_{\ast }},
\label{p}
\end{equation}
\ \ where $\mathcal{V}$\ is the ratio of the $\mathbf{E}\times \mathbf{B}$\
and diamagnetic velocities. This show that $\bar{\rho}$ and 
$\tau_{z}$ (actually $K_{d})$ cannot be modified independently
for ITG turbulence. Their product is $p,$ a dimensionless parameter that
depends on the amplitude of the turbulence (through\ $\mathcal{V}),$ on its
normalized poloidal and parallel correlation lengths and on the gradient
length of the ion temperature. Taking the typical parameters of the ITG
instabilities for ITER size plasmas similar to those of the present tokamaks
(as obtained in numerical simulations), one finds $p\cong 30\left( \rho
_{ion}/\lambda _{y}\right) ^{2}$ (for $R/L_{T_{i}}\cong 5,$ $\lambda _{z}=2\pi
R$\ and $\mathcal{V}\cong 1).$\ 

The dependence of the diffusion coefficient on $\alpha $ particle energy is
shown in Figure \ref{fig8} for several values of $p$. One can see an important
difference between these results\ and those in Figure \ref{fig7}a, which shows that
the decrease of the energy (of $\bar{\rho}$)\ determines a
monotonous increase of the diffusion coefficient. A maximum radial diffusion
appears in Figure \ref{fig8}, which has the amplitude and the location dependent on
the parameter $p.$ The shape of these curves is determined by the
simultaneous variation of $\bar{\rho}$ and $\tau_{z}$ with the energy. The decrease of the energy determines the increase of 
$\tau_{z},$\ which moves from the small decorrelation time
regime in Figure \ref{fig7}c toward the maximum and further to the nonlinear regime
with decaying $D_{1}^{\infty }(K_{d}).$\ The maxima of the curves in Figure
\ref{fig8} correspond to the maximum of $D_{1}^{\infty }(K_{d})$\ in Figure \ref{fig7}c.
This maximum is a decreasing function of the energy, and this is reflected
in Figure \ref{fig8}, which shows that the maximum is smaller when it appears at
larger energies. The diffusion coefficient in Figure \ref{fig8} is small for all the
range of $\alpha $ particle energy at $p=40,$ but at smaller values of $p$\
it can reach significant values (comparable to the ion diffusion
coefficient) at energies much larger than plasma ion energy (see the curve
for $p=10).$ \ \ \ 

Thus, $\alpha $ particle turbulent transport strongly depends on the
parameter $p,$\ which is determined by the characteristics of the
turbulence. Using typical values for the physical quantities in Eq. (\ref{p}%
), the main contribution appears to be determined by the poloidal
correlation length of the turbulence $\overline{\lambda }_{y}=\lambda
_{y}/\rho _{i}.$\ A maximum diffusion coefficient $D_{1}^{\infty }\cong 0.5$
appears in Figure \ref{fig8} for $\alpha $ particle energy of 100KeV if $\overline{%
\lambda }_{y}\cong 2.$\ \ \ \ 

\section{Conclusions}\label{conclusions}

A detailed study of the turbulent transport of the fast $\alpha $ particles
was performed based on the development of the DTM. A realistic model of the
turbulence was considered. It was necessary to determine numerically the
gyro-averaged Eulerian correlation (EC) and to adapt the DTM code to the
discretized EC. The code calculates the time dependent diffusion coefficient
from the EC and its derivatives represented on a space mesh, using an
interpolation procedure.

The dependence of the diffusion coefficient on the five dimensionless
parameters of the model ($K_d, V_d, k_0, a, \bar{\rho}$) was determined and analyzed.

The special shape of the spectrum of the ITG turbulence leads to the decay
of the diffusion coefficient as $1/\bar{\rho}$\ for $\bar{\rho}\gtrsim 2$
for both quasilinear and nonliniar regimes. The decay in the nonlinear
regime is faster than in the case of monotonically decreasing Eulerian
correlations, as is the one considered in \cite{hauff2006turbulent}. The difference between the
quasilinear and the nonlinear regime is given by a factor that depends on the
decorrelation parameter $K_{d}$ and is smaller at large $K_{d}$ in the
nonlinear regime.

The parallel motion of the $\alpha $\ particles provides the main
decorrelation mechanism in ITG turbulence. The characteristic time for this
motion $\tau _{z}=\lambda _{z}/u_{z}$ depends on $\alpha $
particle energy. It is very small when  $\alpha $ particles are born in the nuclear fusion
reaction, and it increases by a factor of the order $20$\ during the cooling
process.\ The dependence of the asymptotic diffusion coefficient on the
energy of the $\alpha $ particles was obtained taking into account both the
parallel motion and the variation of the Larmor radius. The combined action
of these effects leads to the existence of a maximum diffusion coefficient
(Figure \ref{fig8}). We have identified a parameter $p$ (\ref{p}), which includes the
characteristics of the turbulence and the plasma size. The maximum turbulent loss
rate and the corresponding energy are functions of $p.$ Depending on the
specific values of $p$ in a range that is relevant for JET and ITER plasmas,
the turbulent transport of the cooling $\alpha $\ particles can be
negligible or significant. The significant transport appears only when most
of the energy of the $\alpha $\ particles is lost and their energy is in the
range of $100~KeV$ (see Figure \ref{fig8}). 

\ \ \ 

\textbf{Acknowledgements}

This work was supported by the Romanian Ministry of National Education under
the contract 1EU-10 in the Programme of Complementary Research in Fusion.
The views presented here do not necessarily represent those of the European
Commission.

\bibliographystyle{apsrev4-1}
\bibliography{drift}

\end{document}